\documentclass[superscriptaddress,aps,journal=prl,preprint]{revtex4-2}
\usepackage{graphicx}
\usepackage{epstopdf}
\usepackage[utf8]{inputenc}
\usepackage[colorlinks=true, linkcolor=blue, citecolor=blue, urlcolor=blue]{hyperref}

\usepackage{bm,amsmath,amssymb} 
\usepackage{gensymb}
\usepackage{siunitx}
\usepackage{color, soul} 
\usepackage{dcolumn}   
\usepackage{multirow}

\begin{document}

\title{Atomistic Structure of Transient Switching States\\ in Ferroelectric AlScN}

\author{Jiawei Huang}
\thanks{These three authors contributed equally}
\affiliation{Department of Physics, The Hong Kong University of Science and Technology, Clear Water Bay, Hong Kong, China}
\author{Jinyang Li}
\thanks{These three authors contributed equally}
\affiliation{School of Materials Science and Engineering, Harbin Institute of Technology, Shenzhen, Guangdong, 518055, China}
\author{Xinyue Guo}
\thanks{These three authors contributed equally}
\affiliation{School of Materials Science and Engineering, Tsinghua University, Beijing, 100084, China}
\author{Tongqi Wen}
\author{David J. Srolovitz}
\email{srol@hku.hk}
\affiliation{Department of Mechanical Engineering, The University of Hong Kong, Hong Kong, China}
\author{Zhen Chen}
\email{zhen.chen@iphy.ac.cn}
\affiliation{Beijing National Laboratory for Condensed Matter Physics, Institute of Physics, Chinese Academy of Sciences, Beijing 100190, China}
\affiliation{School of Physical Sciences, University of Chinese Academy of Sciences, Beijing 100049, China}
\author{Zuhuang Chen}
\email{zuhuang@hit.edu.cn}
\affiliation{School of Materials Science and Engineering, Harbin Institute of Technology, Shenzhen, Guangdong, 518055, China}
\author{Shi Liu}
\email{liushi@westlake.edu.cn}
\affiliation{Department of Physics, School of Science, Westlake University, Hangzhou 310030, China}
\affiliation{Institute of Natural Sciences, Westlake Institute for Advanced Study, Hangzhou 310024, China}


\begin{abstract}{
We resolve the microscopic mechanism of polarization switching in wurtzite ferroelectric AlScN by integrating advanced thin-film fabrication, ferroelectric switching dynamics characterizations, high-resolution scanning transmission electron microscopy (STEM), and large-scale molecular dynamics simulations enabled by a deep neural network-based interatomic potential. Contrary to earlier interpretations proposing a transient nonpolar intermediate phase, we demonstrate that the broad transitional regions previously observed in STEM images are projection artifacts resulting from the intrinsic three-dimensional zigzag morphology of 180\degree~domain walls, which are a characteristic form of inversion domain boundary. This is further confirmed by STEM imaging of strategically prepared, partially switched Al$_{0.75}$Sc$_{0.25}$N thin films.
Our simulations reveal that switching proceeds through collective, column-by-column atomic displacements, directly explaining the emergence of zigzag-shaped domain walls, and is consistent with the nucleation-limited switching behavior observed in experimental switching dynamic measurements. Furthermore, we show that increasing Sc content systematically lowers domain wall energy and associated nucleation barrier, thereby reducing the switching field in agreement with experimental trends. These findings establish a direct connection between local domain wall structure, switching kinetics, and macroscopic ferroelectric behavior.
}


\end{abstract}

\maketitle

\newpage
The discovery of ferroelectricity in wurtzite-structured Sc-doped AlN (Al$_{1-x}$Sc$_{x}$N, $0.2<x<0.5$) in 2019~\cite{Fichtner19p114103} has introduced a promising materials platform for non-volatile memory technology.
The AlN-based ferroelectrics benefit from the industrial maturity of piezoelectric AlN, which has been widely used for decades in applications such as 4G/5G mobile acoustic wave filters~\cite{Su20p132903,Wang14p133502,Fei18p146,Pinto22p500}. Leveraging this established infrastructure, researchers have developed optimized physical vapor deposition (PVD) processes at sub-$400\degree$C temperatures that are compatible with back-end-of-line (BEOL) semiconductor processing~\cite{Liu21p3753, Motoki24p135105}. A key breakthrough in enabling ferroelectricity in this system is the incorporation of sufficient Sc dopants into the parent wurtzite AlN lattice, which lowers the energy barrier for polarization switching~\cite{Fichtner19p114103, Wang21p104101}. This compositional tuning transforms Al$_{1-x}$Sc$_{x}$N into a ferroelectric, featuring a wide bandgap ($E_g\approx3$--6 eV)~\cite{Deng13p112103}, high remnant polarization ($P_r$) of $>70$--$135~\mu$C/cm$^{2}$~\cite{Fichtner19p114103}, and excellent thermal stability~\cite{Islam21p232905,Wang22p2200005}. Furthermore, the material exhibits thickness-independent $P_r$ below 10--20 nm scale~\cite{Mizutani21p105501,Song25p19491}, overcoming a major scaling limitation in conventional ferroelectric memory technologies.

Despite these advances, a significant challenge persists: the high coercive fields ($E_c \approx 2$–$6$~MV/cm) required for polarization reversal approach the dielectric breakdown threshold ($E_{bd} \approx 6$–$8$~MV/cm), creating a trade-off between switching reliability and dielectric integrity~\cite{Fichtner20p1,Zheng21p144101,Tsai22pSJ1005}. This narrow operating window raises reliability concerns, including high leakage currents and limited endurance. For example, catastrophic leakage currents exceeding $10^{3}$~A/cm$^2$ have been reported at $x = 0.3$~\cite{Yang23p132903}, and endurance failure occurs below $10^5$ cycles~\cite{Mikolajick21p100901}, both orders of magnitude worse than those seen in perovskite and HfO$_2$-based ferroelectrics~\cite{Cheema20p478,Cheema22p65}.
Current mitigation efforts include doping and strain engineering~\cite{Talley18p063802,Yasuoka20p114103,Yazawa21p162903,Yazawa22p042902}. Empirically, the coercive field in Al$_{1-x}$Sc$_x$N decreases with Sc concentration, following $E_c(x) = -15x + 8.35$~MV/cm for $0 < x < 0.43$~\cite{Tsai21p082902}. Additionally, epitaxial tensile strain engineering has been shown to reduce $E_c$ by approximately 30\% at 0.8\% strain~\cite{Schnweger22p2109632}. 

Nevertheless, these approaches currently lack atomistic design principles due to an unresolved debate over the underlying polarization-switching mechanism. Recent \textit{in situ} scanning transmission electron microscopy (STEM) studies have proposed a transient, low-barrier nonpolar intermediate state, characterized by antipolar arrangements of wurtzite motifs when viewed along the [100] crystallographic direction~\cite{Calderon23p1034}.
However, more recent theoretical and experimental work suggests that switching proceeds through conventional inversion domain boundary (IDB*)-mediated mechanisms in the wurtzite lattice~\cite{Wang25p76,wolff2024demonstration,lu2024domain,guido2025ferroelectric,Behrendt25arxivp18816,wolff2025electric}. This apparent discrepancy could stem from the limitations of conventional characterization techniques, which can introduce projection artifacts when imaging three-dimensional (3D) domain structures, as well as from unit-cell-scale simulations that are insufficient to capture the complexity of large-scale domain dynamics.

By integrating multiscale theoretical modeling with advanced thin-film fabrication and STEM imaging, this work provides a comprehensive framework for understanding polarization switching in wurtzite ferroelectrics. Using a deep neural network-based interatomic potential trained on density functional theory (DFT) data, we perform large-scale molecular dynamics (MD) simulations that capture the electric-field-driven 3D evolution of domain structures and transient states in Sc-doped AlN. These simulations reveal that the 180\degree~domain walls exhibit irregular, non-planar morphology, adopting the conventional IDB* configuration. We demonstrate that the previously proposed ``nonpolar" intermediate state arises from projection artifacts caused by the  superposition of coexisting metal (M)-polar and nitrogen (N)-polar domains, along with 180\degree~domain walls. The theoretical predictions are validated by the strong agreement between our simulated structural model and high-resolution STEM images of partially reversed domain states, which were experimentally achieved through tailored electrical pulse engineering.
Investigation of the domain wall propagation reveals a stepwise, column-by-column switching pathway.  This mechanism provides a natural explanation for the formation of the observed irregular domain morphologies and is consistent with measured switching dynamics. Furthermore, we establish a quantitative relationship between Sc concentration and the reduction in both domain wall energy and nucleation barrier, offering a microscopic explanation for the observed doping-dependent reduction in coercive field.

Experimentally, a 300 nm-thick Al$_{0.75}$Sc$_{0.25}$N thin films was deposited on SiO$_2$/Si(001) substrate by a custom-built direct current reactive magnetron sputtering. Polarization switching was characterized using a conventional Positive-Up-Negative-Down (PUND) pulse sequence on an aixACCT TF3000 analyzer, where switched polarization components ($\pm \Delta P$) were isolated from parasitic currents by differential measurement between switching and non-switching pulses. Switching dynamics in Pt/Al$_{0.75}$Sc$_{0.25}$N/Pt capacitors were further investigated through a modified PUND sequence with variable-amplitude/duration write pulses and paired read pulses to extract field- and time-dependent switched polarization while correcting for capacitive/leakage contributions. 
Experiments were conducted using an aberration-corrected JEM-ARM300F2 transmission electron microscope with the cross-sectional specimen extracted from the area below the top electrode on AlScN films. We applied multislice electron ptychography (MEP) method~\cite{chen2020mixed,chen2021electron,chen2021soft4659690} to investigate three-dimensional atomic structures with a picometer lateral resolution and nanometer resolution along the depth dimension (Supplementary Sect. I).
For large-scale MD simulations, we developed a deep neural network-based potential (Deep Potential)\cite{Zhang18p143001}, trained on 20,000 Al$_{1-x}$Sc$_x$N structures with DFT-computed energies and atomic forces (Supplementary Sect. III). The training set covers a broad range of Sc concentrations ($0 \leq x \leq 1$), biaxial strains ($\pm2$\%), and multiple phases, including wurtzite, hexagonal, and rocksalt. DFT calculations were performed using the Vienna \textit{ab initio} Simulation Package (VASP)~\cite{Kresse96p11169,Kresse96p15}, with the projector augmented-wave method and the Perdew–Burke–Ernzerhof (PBE) exchange-correlation functional~\cite{Perdew96p3865}, a plane-wave cutoff of 400~eV, and a \textit{k}-point spacing of 0.2~\AA$^{-1}$.
The validated potential is then employed in MD simulations using LAMMPS~\cite{Plimpton95p1} with a 11520-atom Al$_{0.73}$Sc$_{0.27}$N supercell under $NPT$ conditions at 300 K and 1.013 kPa. An external electric field was applied via the force method~\cite{umari2002ab}, which imposes a field-induced force $F_i = Z_i^* \cdot \epsilon$ on each ion $i$, where $Z_i^*$ is the Born effective charge tensor. Switching pathways identified in MD were further validated by DFT-based climbing-image nudged elastic band (CI-NEB) calculations, using a force convergence threshold of 0.01 eV/\AA.

\begin{figure}[ht]
\centering
\includegraphics[scale=0.5]{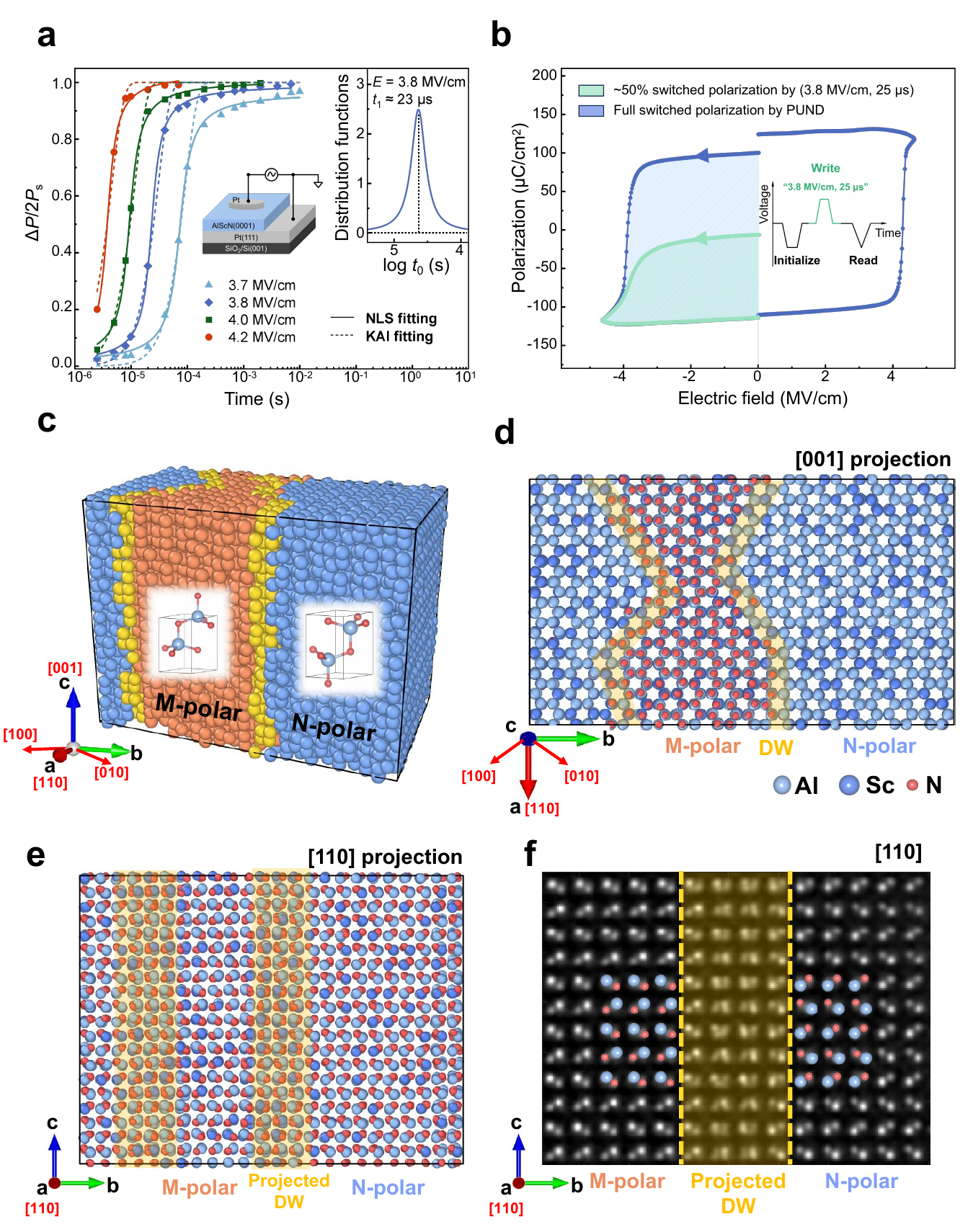}
 \caption{(a) 
 Switchable polarization measured under various external voltages, fitted using both the KAI and NLS models. The NLS model yields a better fitting. The inset shows the Lorentzian distribution function used in the NLS fitting and a schematic of the Pt/AlScN/Pt capacitor structure. (b) Polarization-electric field hysteresis loops demonstrating the fully switched (100\%) and precisely half-switched (50\%) polarization states. The inset illustrates the PUND pulse sequence used to achieve the controlled 50\% switching. (c) Partially switched Al$_{0.73}$Sc$_{0.27}$N supercell obtained with MD simulations. Insets depict local atomic configurations within two oppositely polarized domains, colored in blue and orange. (d) Top view along the [001] direction and (e) front view along the [110] direction of the simulated partially switched structure. (f) High-resolution STEM image showing domain structures in a 50\% switched Al$_{0.75}$Sc$_{0.25}$N thin film. The transitional region between the dashed yellow lines resembles the yellow-shaded area in panel (e).
}
 \label{exp_and_modeling}
 \end{figure}

Our experimental investigation of polarization switching kinetics in Al$_{0.75}$Sc$_{0.25}$N films reveals an important deviation from classical Kolmogorov-Avrami-Ishibashi (KAI) behavior, instead demonstrating nucleation-limited switching (NLS) dynamics (Fig.~\ref{exp_and_modeling}a). This behavior can be attributed to the film’s strong (0002) texture, where grain boundaries act as barriers to domain wall propagation, causing polarization reversal to be dominated by the stochastic nucleation of isolated reversal nuclei (Supplementary Sect. II)~\cite{zhou2024enhanced,guido2024kinetics}. Importantly, the extracted distribution of characteristic switching times (Fig.~\ref{exp_and_modeling}a, inset) provides a quantitative basis for implementing controlled partial switching. Leveraging the NLS-derived temporal statistics, we designed an optimized electric-field pulse (3.8 MV/cm, 25 $\mu$s) precisely timed to coincide with the peak switching probability. Subsequent polarization mapping confirms that this single-pulse protocol reliably produces a metastable , partially reversed state with $\approx50$\% switched polarization (Fig.\ref{exp_and_modeling}b). The exceptional stability of this configuration, stemming from the high intrinsic coercivity of wurtzite ferroelectrics, provides an ideal platform for atomistic structural analysis.

We further use MD simulations to obtain a partially switched state in an Al$_{0.73}$Sc$_{0.27}$N supercell of 11520 atoms. As shown Fig.~\ref{exp_and_modeling}c, the system contains coexisting N-polar (unswitched) and M-polar (switched) domains. From the top view along the $c$-axis ([001] direction), the interface between oppositely polarized domains forms atomically sharp 180\degree~domain walls (highlighted in yellow in Fig.~\ref{exp_and_modeling}d). This boundary adopts the structure of a conventional IDB* (see discussions in below), but unlike the planar wall typical of perovskite ferroelectrics, it exhibits a highly irregular, zigzag morphology. This structural complexity is crucial for interpreting the experimental STEM images.
When viewed along the $a$-axis ([110] direction; see front view in Fig.~\ref{exp_and_modeling}e), the projected structure appears to exhibit broad transitional regions spanning several atomic layers (highlighted in yellow). 
Our high-resolution STEM MEP imaging (Fig.~\ref{exp_and_modeling}f) from an irregular wall of partially switched Al$_{0.73}$Sc$_{0.27}$N samples reveals
atomic configurations (Supplementary Sect. I)
that closely match the simulated projection (Fig.~\ref{exp_and_modeling}e). The strong agreement between MD simulations and STEM imaging suggests that the seemingly broad interfacial regions are not intrinsic features, but rather result from projection effects, supporting the existence of irregular, non-planar 180\degree~domain walls.

\begin{figure}[ht]
\centering
\includegraphics[scale=0.58]{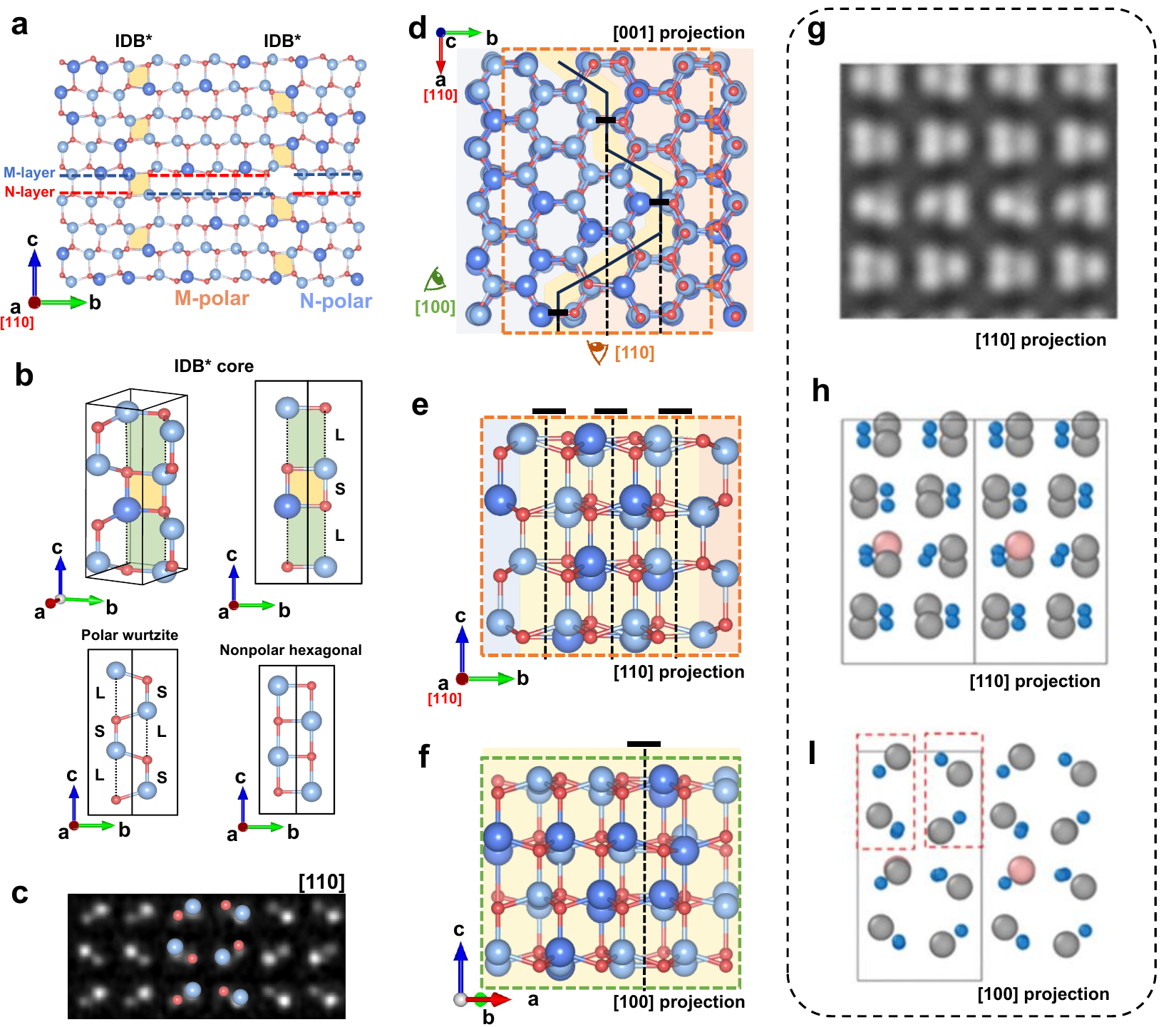}
\caption{
(a) [110] projection of a two-atomic-layer slice with 180\degree~domain walls. Four-membered bond rings in the IDB* configuration are highlighted in yellow. 
(b) 3D schematic of an IDB* core, featuring alternating long (L) and short (S) M--N bonds along the $c$-axis. For comparison, the structures of polar wurtzite AlN and the commonly assumed nonpolar hexagonal phase are shown in the lower panel. 
(c) Experimental STEM image of the IDB* configuration. 
(d) Top view ([001] projection) of the simulated domain structure. M-polar and N-polar regions are colored blue and red, respectively; domain wall regions are shown in yellow. 
(e) [110] projection and (f) [100] projection of the structure shown in (d). 
(g) STEM image reported in ref.~\cite{Calderon23p1034}. 
(h) [110] and (i) [100] projections of the proposed nonpolar intermediate phase deduced from (g).
  }
  \label{transient}
 \end{figure}

To resolve the atomic-scale structure of the 180\degree~domain wall, we analyze a representative two-atomic-layer slice, projected along the [110] direction (Fig.~\ref{transient}a). This configuration exhibits typical features of an IDB*, consistent with Northrup’s model~\cite{Northrup96theory}, including four- and eight-membered bond rings. A three-dimensional view (Fig.~\ref{transient}b) reveals that while the four-membered rings are coplanar within the $bc$-plane, the eight-membered rings are buckled, connecting adjacent atomic layers along the $a$ axis.
At the core of this IDB*, we observe alternating short ($\approx1.9$~\AA) and long ($\approx3.0$~\AA) vertical M-N bonds. The observed bond-length alternation signifies a Peierls-like distortion. In contrast, the undistorted paraelectric hexagonal phase, often regarded as the nonpolar reference state for wurtzite ferroelectrics~\cite{farrer2002properties,speck2009nonpolar,Dreyer16p21038,huang2022origin,yang2025physics}, features equivalent vertical M–N bonds along the $c$-axis (see Fig.~\ref{transient}b).
Importantly, the characteristic IDB* atomic configuration, including the signature bond rings, is directly resolved in our phase image of a middle slice in the MEP reconstructed result 
from one region with straight wall parallel to the projection direction in
a partially switched AlScN film (Fig.~\ref{transient}c), providing strong experimental validation for our simulated structural model.

We now explain how the 3D domain structure with atomically sharp domain walls gives rise to the wide interfacial regions observed in STEM imaging.
The simulated top view along the $c$-axis ([001] projection, Fig.~\ref{transient}d) shows that the domain wall follows an irregular, zigzag path.
Rather than forming a flat, planar interface, the boundary meanders laterally across multiple lattice constants along the $b$-axis. Three thick black lines, each orthogonal to a segment of the domain wall aligned with the $a$-axis, indicate the locations of the three IDB* boundaries depicted in Fig.~\ref{transient}a.
When viewed along the $a$-axis ([110] projection, Fig.~\ref{transient}e), this zigzag geometry causes overlapping of N-polar and M-polar domains, as well as of the domain wall segments themselves. This overlap results in the appearance of a broad, diffuse transitional region (yellow-shaded area), an effect that closely matches both our experimental STEM data (Fig.~\ref{exp_and_modeling}e) and previous observations (reproduced here as Fig.~\ref{transient}g)~\cite{Calderon23p1034}.
The simulated projection along the [100] direction (Fig.~\ref{transient}f) again displays features of the  transitional region. However, this view differs significantly from that of a previously proposed intermediate state (Fig.~\ref{transient}i), which had been interpreted as evidence for a distinct, nonpolar phase with antipolar arrangements of wurtzite
motifs~\cite{Calderon23p1034}.
Our results demonstrate that the broad transitional regions observed in 2D projections are not indicative of a separate physical phase. Rather, they are projection-induced artifacts arising from the complex 3D domain structure with zigzag atomically sharp domain walls. 

\begin{figure}[ht]
\centering
\includegraphics[scale=0.46]{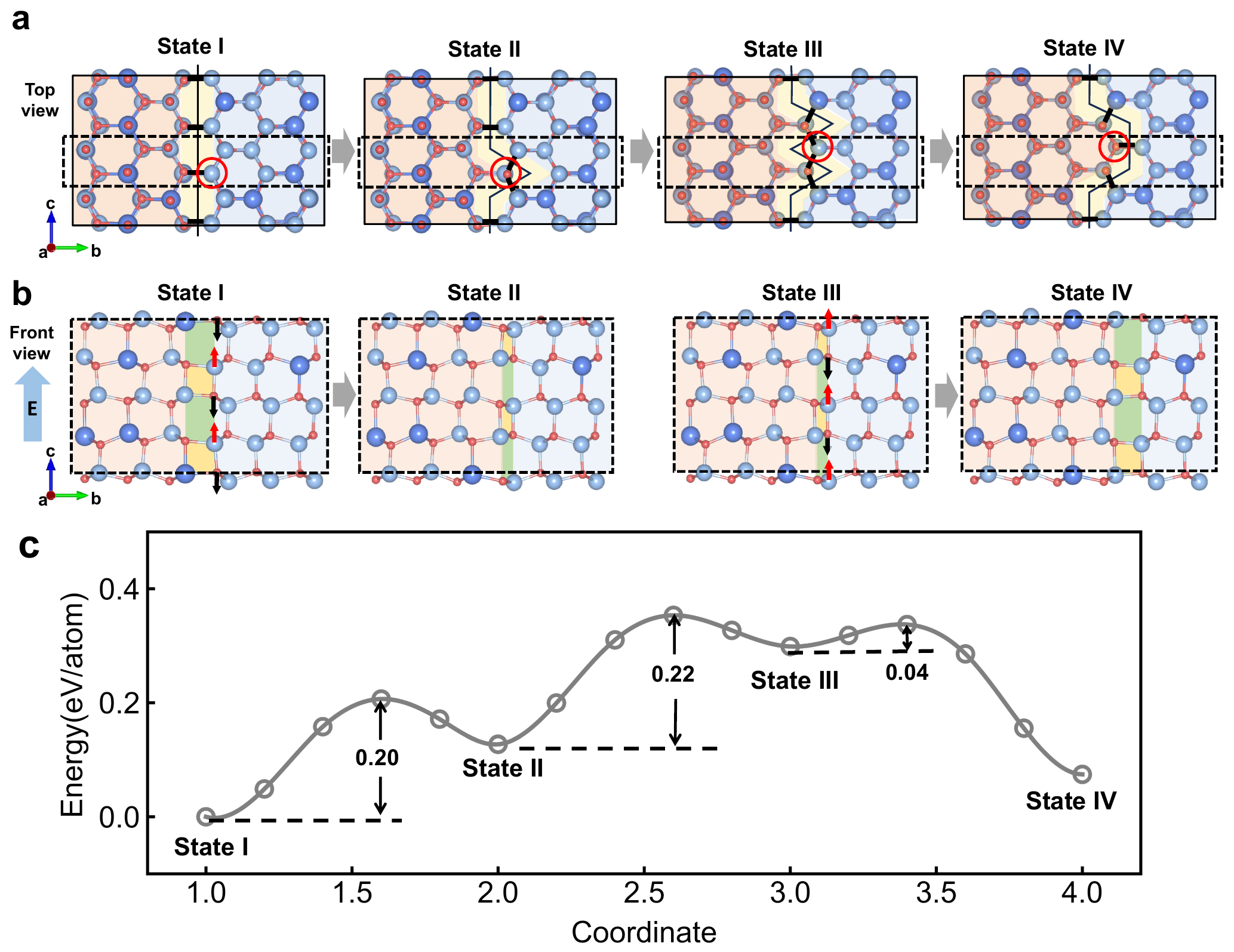}
\caption{
    (a) Key stages of domain wall propagation extracted from MD simulations. Top views show the evolution of the domain wall, starting from a flat interface (State I), progressing through complex intermediate configurations (States II and III), and locally re-flattening in State IV.  
    (b) Corresponding front view of a two-atomic-layer slice, highlighted by the dashed rectangle in (a). 
    (c) Energy landscape associated with domain wall propagation process in (a), calculated using the DFT-based CI-NEB method.
}
  \label{neb}
 \end{figure}

We find from MD simulations that the switching mechanism in Al$_{0.73}$Sc$_{0.27}$N proceeds via localized, collective atomic shifts along the polar axis. As shown in Fig.~\ref{neb}a, a typical process begins with an initial flat domain wall (State I). Under an upward electric field, the reversal of a single N-polar column near the wall transforms the local structure into an M-polar configuration, resulting in a more irregular domain boundary (State II).
The corresponding front view of a two-atomic-layer slice (dashed rectangle in Fig.~\ref{neb}a) shows the field-driven atomic displacements: Al/Sc atoms shift in the direction of the field, while N atoms move in the opposite direction (see State I to State II in Fig.~\ref{neb}b).
The subsequent reversal of adjacent columns further increases the roughness of the domain wall morphology (State III), leading to a rise in interfacial energy. To reduce this energy, the system rapidly evolves into a lower-energy configuration by reestablishing a more locally flat domain boundary (State IV). This relaxation step is again mediated by the reversal of an individual atomic column, as illustrated in the corresponding front view. 

The energy evolution of the entire switching process, obtained from DFT-based NEB calculations, is shown in Fig.~\ref{neb}c. 
We note that the calculated energy barrier per supercell is $\sim$0.8eV. In previous studies, the barrier was often normalized by dividing this value by the total number of atoms in the supercell~\cite{liu2023doping,bhattarai2024effect,hwang2024first}. However, since only a single column of atoms is switched in each step, this approach may not accurately reflect the switching mechanism. Therefore, we introduce a normalized energy barrier, defined as the total energy barrier divided by the number of atoms involved in the column-wise switching process.
The switching of an atomic column at a locally flat domain wall is associated with an normalized energy barrier of approximately 0.2~eV/atom (Supplementary Sect.~IV), as observed in both the State~I~$\rightarrow$~State~II and State~II~$\rightarrow$~State~III transitions. Furthermore, the progressive increase in energy from State~I to State~III indicates that the domain wall energy scales with its roughness. The elevated energy of State~III renders it unstable, and the subsequent relaxation to a geometrically simpler configuration (State~IV) proceeds via a significantly lower energy barrier of just 0.04~eV/atom.
These results offer insight into the microscopic origin of the zigzag domain walls observed in large-scale MD simulations. The switching of individual atomic columns at locally flat domain walls involves similar energy barriers, and their spatial separation allows them to act as largely independent switching units. Consequently, switching occurs stochastically, with different columns reversing at different times under the applied electric field. This asynchronous behavior gives rise to rough, non-planar domain boundaries that manifest as zigzag-shaped walls at larger scales. Moreover, the high energy of irregular configurations like State III drives the system to locally re-flatten the wall, reinforcing the dynamic, sawtooth-like morphology of the domain wall.

\begin{figure}[ht]
\centering
\includegraphics[scale=0.5]{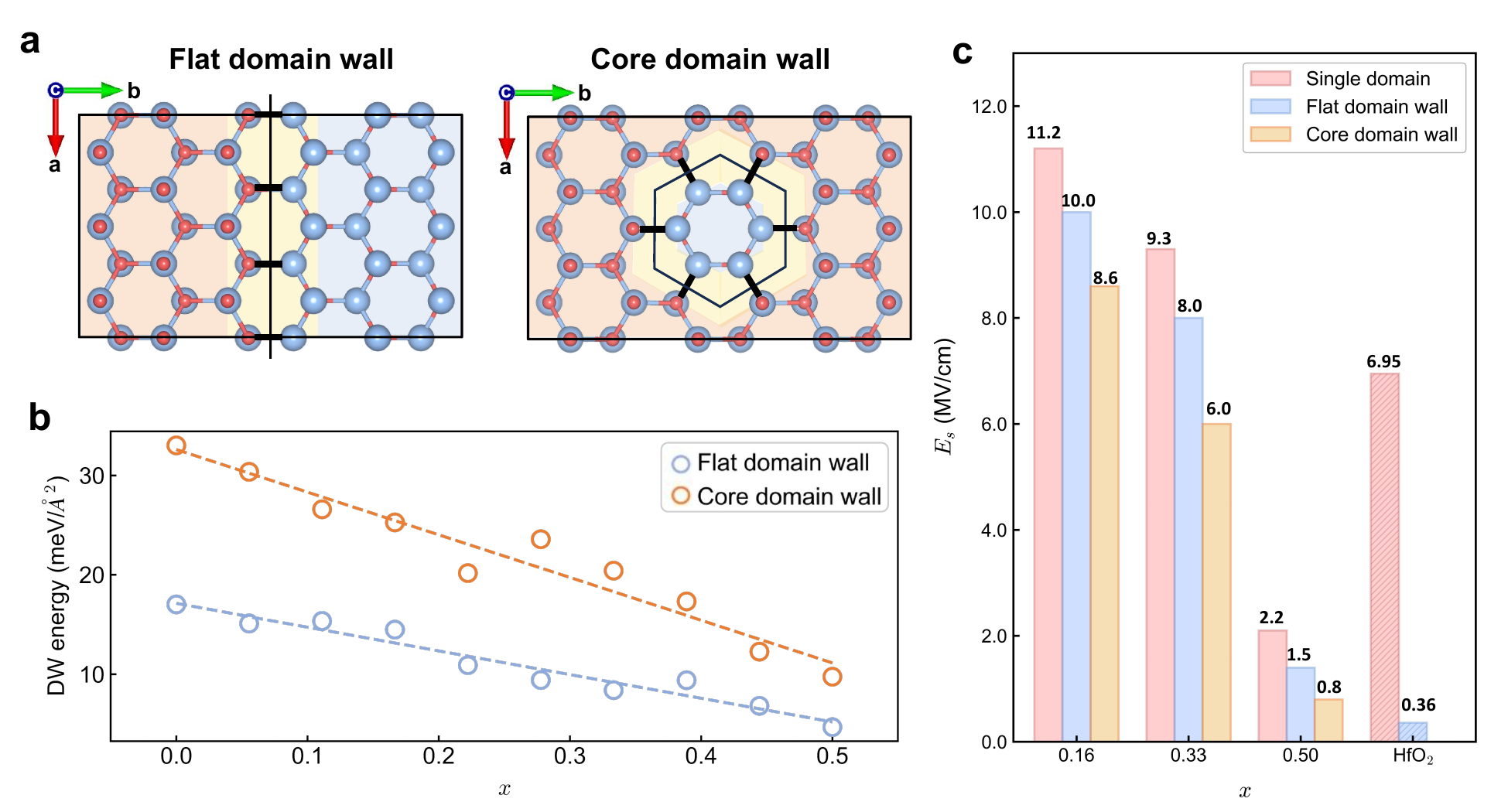}
\caption{    
(a) Schematic of two types of domain walls.
(b) Domain wall (DW) energy as a function of Sc concentration ($x$) for both flat and core domain walls, obtained from MD simulations.  
(c) Switching field ($E_s$) as a function of Sc concentration ($x$) for different domain wall configurations. Reference values of $E_s$ for ferroelectric HfO$_2$ in single-domain and domain-wall-containing supercells are included for comparison.
}
  \label{doping}
 \end{figure}

The established link between polarization switching and domain wall motion suggests that the coercive field can be tuned by reducing the domain wall energy through doping. Lowering the domain wall energy not only facilitates wall motion but also decreases the nucleation barrier, as the primary energy cost of nucleation arises from the formation of new domain walls~\cite{liu16p360,yang2025theoretical}.
We investigate the effect of Sc concentration on the energetics of two representative domain wall configurations shown in Fig.~\ref{doping}a: a conventional flat domain wall and a core domain wall, which forms at the interface of a hexagonal, six-atomic-column N-polar nucleus embedded within an M-polar matrix. This quasi-one-dimensional reversed domain, identified in our MD simulations, represents the minimal stable switching unit within a parent domain and may serve as an effective nucleation center.
As shown in Fig.~\ref{doping}b, the formation energies of both types decrease nearly linearly with increasing Sc content. The flat wall energy drops from 17.1 to 4.6 meV/\AA$^2$, while the core domain energy decreases more significantly, from 33.1 to 9.7meV/\AA$^2$. 
This trend continues up to the critical doping threshold ($x \approx 0.5$), beyond which the material becomes a non-ferroelectric hexagonal phase. 

We assess the ease of polarization switching by determining the minimum field strength ($E_s$) required to induce switching within 100 ps in MD simulations, following a protocol developed in ref.~\cite{zhu2025origin}. This evaluation is performed for single-domain supercells, as well as for systems containing flat and core domain walls.
Figure~\ref{doping}c reveals several key findings. First, increasing Sc concentration reduces the magnitude of $E_s$, consistent with experimental observations. This effect is attributed to the doping-induced reduction in domain wall formation energy, which lowers the barriers for both domain wall motion and nucleation.
Second, pre-existing domain walls further reduce $E_s$ compared to the single-domain case. However, this domain-wall-assisted switching is less pronounced in AlScN. For example, at $x = 0.33$, flat domain walls only moderately reduce the switching field from 9.3 MV/cm to 8.0 MV/cm. In contrast, ferroelectric HfO$_2$ shows a much larger reduction, nearly an order of magnitude, due to domain wall presence, highlighting the nucleation-limited nature of switching in AlScN.
Lastly, core domain walls are more effective than flat ones in lowering the coercive field, suggesting a potential design strategy for reducing switching fields in AlScN systems. 

In summary, by integrating multiscale simulations with experimental validation, we resolve a central debate regarding the polarization switching mechanism in wurtzite ferroelectrics. The broad transitional regions previously observed in STEM images are identified as projection artifacts arising from the intrinsic three-dimensional zigzag morphology of 180\degree~ domain walls, which are a form of inversion domain boundary.
Our analysis reveals that switching proceeds via a highly localized, column-by-column reversal process, consistent with a nucleation-limited switching model observed in experiments. The zigzag shape of the domain walls originates from asynchronous switching of individual atomic columns, which roughens the wall morphology under an applied electric field.
Moreover, the experimentally observed reduction in coercive field with increasing Sc doping is attributed to a doping-induced decrease in domain wall energy, which lowers the energy barriers for both domain wall motion and nucleation.
Together, these findings provide a comprehensive, atomic-scale understanding of ferroelectric switching in AlScN and offer useful insights for the design and optimization of wurtzite-based ferroelectric materials.

\begin{acknowledgments}
S.L. acknowledge the supports from Zhejiang Provincial Natural Science Foundation of China (LR25A040004).
The computational resource is provided by Westlake HPC Center. Z.H.C. acknowledges the financial support from the National Natural Science Foundation of
China (Grant Nos. 52525209, 92477129 and 52372105) and Guangdong Basic and Applied Basic Research Foundation (Grant No. 2024B1515120010). Z.C. acknowledges the financial support from the National Natural Science Foundation of China (Grant No. 52273227).
\end{acknowledgments}
\newpage

\bibliography{SL}

\newpage

 \newpage

\newpage

 \newpage


\end{document}